# Tuning the dynamics of magnetic droplet solitons using dipolar interactions

H. F. Yazdi, [1] G. Ghasemi, [1] Majid Mohseni, [1,*] and Morteza Mohseni [2]

[1]Department of Physics, Shahid Beheshti University, Tehran 19839, Iran

[2] Fachbereich Physik and Landesforschungszentrum OPTIMAS, Technische Universität Kaiserslautern, 67663 Kaiserslautern, Germany

Magnetic droplets are dissipative magnetodynamical solitons that can form under current driven nanocontacts in magnetic layers with large perpendicular magnetic anisotropy. Here, we extend the original droplet theory by studying the impact of the dipolar interactions on the dynamics of droplet solitons. By varying the thickness of the free layer of a spin torque nano-oscillator, we systematically tune the internal field of the free layer to investigate the dynamics of droplet solitons. Our numerical results show that increasing the free layer thickness increases the droplet's threshold current, decreases the droplet's frequency and diameter, enlarges the current hysteresis and also modifies the structure of the droplet. The Oersted field of the current breaks the precessional phase coherency and deteriorates the stability of the droplet in free layers with larger thicknesses. Moreover, our findings show a simple relation to determine the impact of the free layer thickness on the droplet's nucleation boundaries. Our study presents the missing brick on the physics behind magnetic droplet solitons, and further illustrates that magnetic droplets in thinner layers possess more promising characteristics for spintronic applications and enable devices with higher speed of operation.

## I. Introduction

Spin-waves (SWs) are the collective eigen-excitations of a magnetic order parameter [1]. The use of SWs provides new avenues to design and implement data processing units relying on the wave-based and neuromorphic computing architectures [2,3]. SWs in ferromagnetic systems without chirality possess an anisotropic dispersion, i.e. the shape of the dispersion depends on the relative orientation of the magnetization vector and the SW wave vector [2,4]. This is mainly caused by the impact of the dipolar interactions on the SW dynamics, as well as the influence of the exchange energy in high wave vector SWs. Unlike the exchange interactions, the dipole-dipole interaction scales with distance, and therefore, its contributions to the SW dynamics can be safely neglected in sufficiently ultra-thin films [5-7].

The abundant nonlinearity of SWs provides rich opportunities to investigate the basics of nonlinear wave dynamics. Examples can be nonlinear multi-magnon scattering processes [8-11], parallel parametric amplification [12,13], auto-oscillation based on the spin-transfer torque (STT) effect [14,15] and soliton formation [16,17]. In particular, using the STT to drive the magnetization oscillations allows the investigation of nonlinear SW dynamics on the nanoscale, and to design device architectures for post-von Neumann computing [3]. More interestingly, it allows for the nucleation of a large variety of solitons and spin textures, for example, skyrmions [18], magnetic bullets, and droplets [17,19-21].

Magnetic droplet solitons (droplet hereafter) are strongly nonlinear and localized SW solitons that can form in thin magnetic layers with sufficiently large perpendicular magnetic anisotropy (PMA). To nucleate a droplet, it is necessary to overcome the viscous magnetic damping of the system by utilizing the STT. This can be achieved using a nanocontact (NC) geometry in spin-torque nano-oscillators (STNOs) [19,20,22]. In such devices, the injected dc into the NC become spin-polarized due to the spin-dependent conductance at the interfaces. The flow of the spin-polarized electrons transfers spins angular momentum to the second ferromagnet, the free layer in which the droplet nucleates [21]. This essentially means that droplets are dissipative solitons since they only sustain if the damping of the system is compensated by the STT (gain). The strong nonlinearity of the droplets given by their very large angle of precession leads to various intriguing nonlinear and instability dynamics [23-31]. This also leads to a strong output power compared to similar devices, suggesting the benefits of droplet-STNOs in microwave data processing devices, magnonic and spintronic applications.

In addition to the gain-loss balancing, the nucleation of the droplet depends on the counteraction of the PMA against the exchange energy. Indeed, the original theory of the droplet considers an ultrathin magnetic layer in which the SW dispersion is dominated by the exchange interactions [21]. This theory well explains the observed dynamics of droplets in ultrathin Ni/Co multilayers whose thicknesses do not exceed more than a few nanometers and therefore, the influence of the dipolar interactions are negligible [19,20]. This assumption is valid as long as the thickness of the free layer is much smaller than the SW wavelengths [21]. However, this assumption prevents an explicit understanding of the role of the dipolar interactions on the dynamics of droplets.

Moreover, recent progresses on the growth of low damping thin films with large PMA featuring larger thicknesses compared to Ni/Co multilayers [32], as well as the ability of dipolar fields to tune the chirality and stabilize spin textures [33] motivate us to investigate the role of dipolar interactions on the dynamics of droplet solitons. However, conducting analytical calculations for this aim in such a strongly nonlinear and nonuniform magnetodynamics is a rather tedious and challenging task that encourages the use of numerical approaches to tackle this issue. Indeed, numerical simulations based on the micromagnetic frameworks is now a reliable and well established method to study magnetodynamics in nanostructures, including various nonlinear dynamics of droplets [26,29,30,34-37]. Needles to mention that the studies based on such methods turned out with excellent agreements to the experimental observations and theoretical analysis.

Here, we use micromagnetic simulations to investigate the role of dipolar interactions on the dynamics of magnetic droplet solitons. To fulfill this task, we tune the free layer thickness of a STNO with large PMA. Using this approach enables us to systematically tune the internal field of the free layer only due to the dipolar interactions, while it keeps the rest of the parameters unchanged. It will be shown how increasing the thickness influences the threshold current, frequency, shape and the diameter of the droplet. Moreover, stability and hysteresis of the droplet as a function of the thickness will be discussed. Finally, the role of the dipolar fields on the droplet nucleation boundaries will be presented.

## I. Methods

The Landau-Lifshitz-Gilbert-Slonczewski equation which describes current-driven motion of magnetization in a magnetic layer has the form [21],

$$d\bm{m}/dt = -\bm{m} \times \bm{h}_{\text{eff}} - \alpha \bm{m} \times (\bm{m} \times \bm{h}_{\text{eff}}) + \sigma \bm{m} \times (\bm{m} \times \bm{m}_{\text{f}}). \qquad (1)$$

in which $\bm{m} = \bm{m}(\bm{r}, t) = \bm{M}/M_s$ is the unit magnetization vector and $M_s$ the saturation magnetization. The first term on the right-hand side describes the magnetization precession in an effective field $\bm{h}_{\text{eff}}$ that reads,

$$\bm{h}_{\text{eff}} = \bm{h}_{\text{demag}} + \bm{h}_{\text{exch}} + \bm{h}_{\text{ani}} + \bm{h}_{\text{ext}} \qquad (2)$$

which includes the standard contributions of the demagnetizing field due to the magnetostatic energy, exchange, perpendicular uniaxial magnetic anisotropy and the external bias field. The second term on the right-hand side of the Eq. (1) describes the Gilbert damping with $\alpha$ being the dimensionless Gilbert damping parameter. This damping is compensated by the third terms which describes the STT whose σ is the parameter that defines the magnitude of the driving spin polarized current. Moreover, $\bm{m}_{\text{f}}$ is a unit vector defining the direction of the spin polarization of the driving current, which is characterized by the fixed layer magnetization vector.

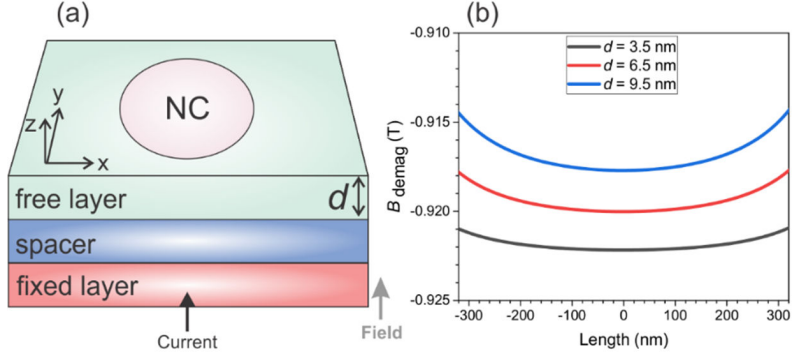

FIG.1. (a) Schematic picture of the device under investigation. The spin torque nano-oscillator consists of a fixed and a free layer which are separated by a spacer. The thickness of the free layer which is simulated is varied. (b) Spatial distribution of the demagnetizing field inside the free layer while the thickness is varied.

The system under investigations is schematically shown in Fig. 1(a). We use MuMax 3.0 open source package to simulate the space- and time-dependent magnetization dynamics in a free layer of a STNO [38]. The system has a dimensions of 1000 nm × 1000 nm × $d$ (nm), in which $d$ represents the free layer thickness and varies systematically between 3.5 nm to 9.5 nm. The system is divided into 512 × 512 × 1 cells. The following realistic parameters of the Ni/Co multilayers in which the droplet has been observed and its dynamics are well investigated are used in the simulations [19]: saturation magnetization $M_s = 737 kA/m$, Gilbert damping α = 0.03, perpendicular magnetic anisotropy $K_u = 434\ kJ/m3$, and the exchange stiffness $A_{exch} = 15\ pJ/m$. The NC diameter is fixed to 100 nm, and the spin polarization assumed to be out-of-plane with spin torque asymmetry of $\lambda = 1.3$ and polarization equal to $P = 0.43$. We neglect the Oersted field of the current unless mentioned otherwise. This prevents the onset of the drift instability processes of the droplet. In addition, this allows a simplified modeling and the possibility to compare the droplet dynamics in layers with different thicknesses. The external field is applied normal to the plane. All simulations have been carried out in the absence of thermal noise, and the polarity of the continuous dc which is used to drive the magnetization oscillations is set to negative.

In order to study the magnetization dynamics, the dynamical components of the magnetization $m_x$(x,y,t) or $m_z$(x,y,t) are collected for the duration of 20 ns starting from the nucleation time of the droplet. The intensities of the excited modes were calculated by performing a fast Fourier transformation (FFT) on the collected data. The nucleation time of the droplet is defined as the moment when the magnetization below the NC reverses, which was found by analyzing the time trace of the out-of-plane component of the magnetization below the NC [37].

We first set the external field to μ₀$H_e$ = 0.4 T. In order to show the impact of the thickness on the internal field of the free layer, we present the spatial distribution of the out-of-plane component of the demagnetizing field inside the free layer with various thicknesses in Fig. 1(b). Evidently, increasing the thickness from $d$ = 3.5 nm to $d$ = 9.5 nm leads to a smaller absolute value of the demagnetizing field $B_{demag}$ inside the free layer. Note that such a variation of the demagnetizing field directly increases the effective field below the NC in accordance with Eq. (2). In addition, it is expected that increasing the thickness leads to a larger contribution of the dipolar interactions to the SW dynamics [7,12]. In the following, we systematically study how this change of the internal field influences the dynamics of the droplet solitons.

# I. Results and discussions
## A. Droplet nucleation and high frequency dynamics

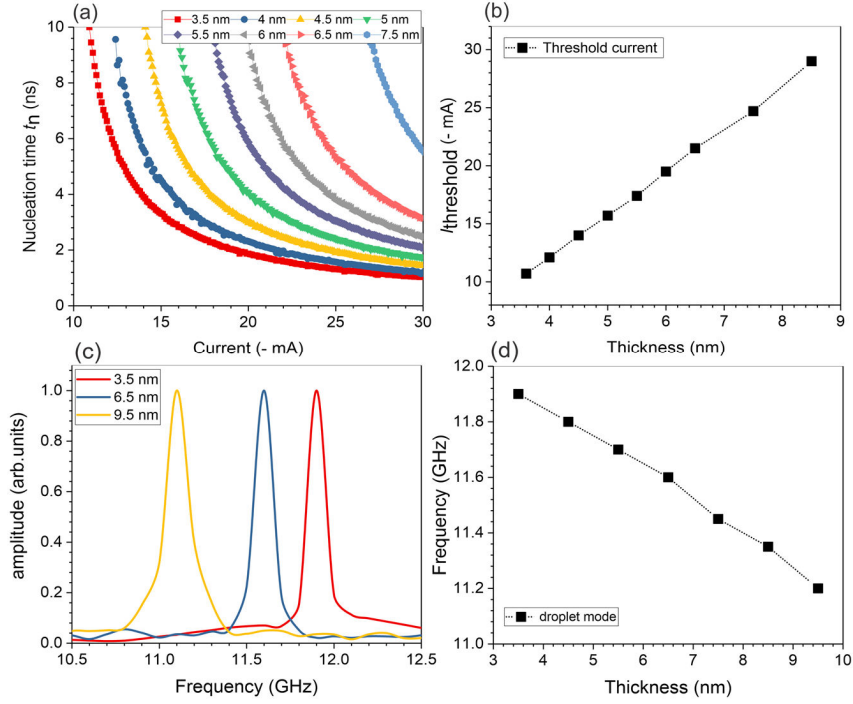

FIG.2. (a) Droplet nucleation time with respect to the applied current in free layers with different thicknesses. (b) The droplet threshold current with respect to the thickness of the free layer which is deduced from (a). (c) Frequency spectra of the droplet mode in free layers with different thicknesses. (d) The frequency of the droplet mode with respect to the thickness of the free layer.

We begin our discussions by analyzing the threshold current $I_{threshold}$ and the nucleation time $t_n$ of the droplet. In principle, the applied current density must be high enough to compensate the damping of the free layer to permit the nucleation of a droplet. This current density possesses a clear threshold which is higher compared to the auto-oscillation mode (Slonczewski solution) due to the strong nonlinearity of the droplet [19,21]. Moreover, such a nonlinear dynamic requires a short time to develop in the system, named as the nucleation time $t_n$ [27,37]. Figure 2(a) presents the nucleation time of the droplet $t_n$ with respect to the applied current to the NC, while the thickness of the free layer varies. It is evident that applying a higher current into the NC allows the droplet to nucleate faster, given by the faster compensation of the damping. This process is independent of the free layer thickness. The nucleation time approaches to sub nanoseconds in the presence of sufficiently high currents. Although it will never become equal to zero. In addition, one can see that increasing the thickness of the free layer has a direct impact on the threshold current $I_{threshold}$ of the droplet.

To better compare the obtained results in different devices, in Fig. 2(b) we present the thickness dependence threshold current of the droplet that is deduced from Fig. 2(a). We observe a linear dependence of the threshold current on the free layer thickness. The results presented in Fig. 2(a-b) demonstrate an additional degree of influence on the determination of the droplet threshold current in comparison to the original theory [21]. We would like to mention that the obtained results are independent of the sample geometry, unless the nonuniformity of the demagnetizing field is strong enough to influence the dynamics, e.g. in a nanowire or if the NC is placed near the edges of the setup. However, the latter configuration can lead to drift instabilities and droplet propagation [25,30].

The droplet is a magnetodynamical soliton featuring a precessional frequency in the GHz range [19]. Indeed, the droplet core reverses with respect to the bias field, while the droplet boundaries precess with a frequency between the Zeeman frequency and the spatially uniform precession mode, the Ferromagnetic Resonance (FMR) frequency of the system [21]. Figure 2(c) represents the frequency of the droplet mode within free layers of different thicknesses that are equal to 3.5 nm, 6.5 nm and 9.5 nm. We fixed the current to $I = 45$ mA in order to remain consistent from one device to the other. This permits us to fix the externally injected energy into the system and to exclude the influence of the current density on the frequency of the droplet [19-21]. One can observe that in free layers with higher thicknesses: (*i*) the droplet frequency decreases from $f = 11.9$ GHz (3.5 nm) to $f = 11.2$ GHz (9.5 nm) and, (*ii*) the droplet linewidth increases from $\Delta f = 78$ MHz (3.5 nm) to $\Delta f = 135$ MHz (9.5 nm). Indeed, the droplet frequency decreases linearly with respect to the free layer thickness as displayed in Fig. 2(d). Such a trend suggests that the droplet will turn into a static or low frequency topologically trivial magnetic bubble in very thick films [21], e.g. larger than 120 nm in our setup. However, driving the magnetization of such thick layers using STT is extremely inefficient and moreover, the droplet threshold current exceeds reasonable values that can be tolerated by devices. The smaller frequencies of the droplet in larger thicknesses cannot be explained by a higher effective field below the NC, however, it is due to the impact of the non-local magnetostatic energy and dipolar interactions in larger thicknesses, which dominate the contributions of exchange energy to the SW dynamics [21]. This finally leads to the stabilization of the droplet as a static bubble.

### B. Droplet annihilation and hysteresis

The droplet soliton has a very narrow condition to form due to its dissipative nature [19-21]. This also leads to the presence of a current hysteretic response in the absence of drift instabilities, regulated by the nucleation and annihilation processes [20-21]. This magnetic bistability feature of the droplet allows the magnetic state below the NC to be dependent on its prior history.

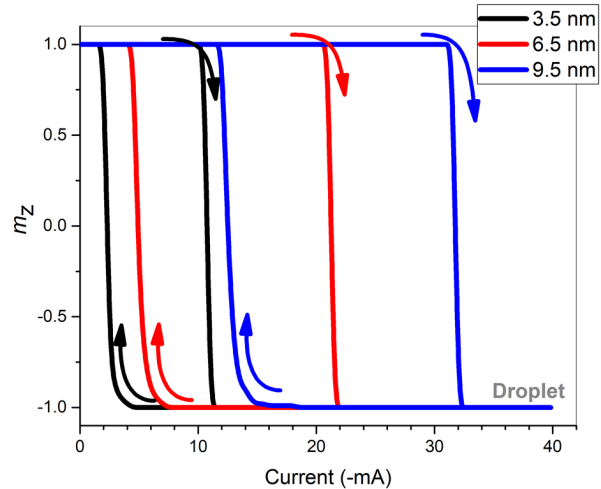

FIG.3. The droplet current hysteresis as in free layers with different thickness

We now investigate the hysteresis characteristics of the droplet as a function of the free layer thickness. To this end, we sweep up the current from $I = 0$ mA to $I = 40$ mA, and then sweep down the current from $I = 40$ mA to $I = 0$ mA. We record the out-of-plane component of the magnetization below the NC and plot it with respect to the applied current. Figure 3 shows the current hysteresis characteristics of the droplet in layers with different thickness. The droplet nucleates ($m_z = -1$) once the applied dc exceeds the threshold of the modulational instability of small amplitude SWs below the NC, as discussed in the preceding section. However, the droplet annihilates ($m_z = 1$) in a given dc is much smaller than current threshold when decreasing the current, which is caused by the fact that the required spin-torque to sustain the droplet is less than of its nucleation. Moreover, increasing the thickness leads to a larger current hysteresis loop which can be explained by the droplet nucleation boundaries and a higher effective field below the NC in free layers with larger thickness [20].

### C. Structure of the droplet and stability

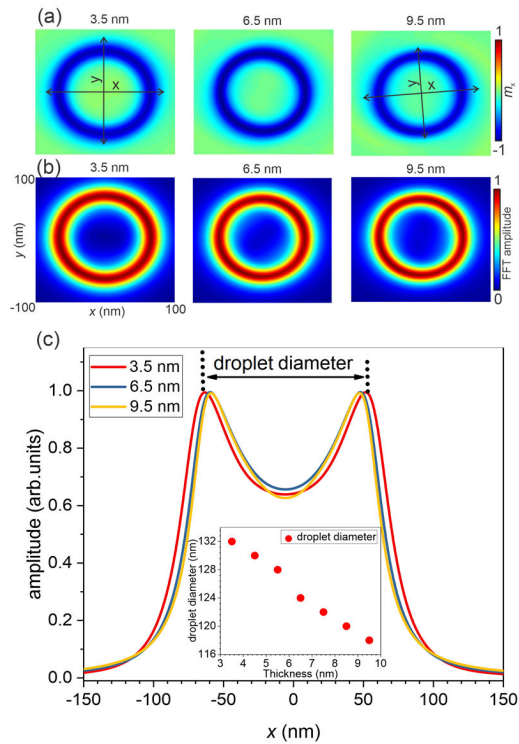

FIG.4. (a) Snapshots of the droplet dynamics in free layers with different thicknesses. (b) The space-time averaged spatial distribution of the droplet mode in free layers with different thicknesses. (c) One-dimensional spatial distributions of the droplet mode extracted from (b) at $y = 0$ in free layers with different thicknesses. The inset shows the droplet diameter with respect to the thickness of the free layer.

We now analyze the structure of the droplet to better understand the impact of the dipolar field on the droplet dynamics and also to investigate the possible drift instabilities of the droplet from the NC [25, 29, 39]. Figure 4(a) displays the snapshots of the of the droplet dynamics within 3.5 nm, 6.5 nm and 9.5 nm thick free layers. Our simulations demonstrate that the droplet precessional boundaries gradually evolves into a semi-elliptical shape via increasing the thickness. This is evidenced by considering the ratio of the precessional axis of the droplet x / y, which is close to unity for the droplet in a 3.5 nm layer and increases to x / y = 1.15 for the droplet in the 9.5 nm thick free later.  We believe that is mainly caused by a larger contribution of the dipole-dipole interaction to the SW dynamics in the system in thicker films, which let the droplet to acquire an anisotropic structure. In other words, due to the anisotropic nature of the dipolar interactions, their higher contributions to the SW dynamics can break the symmetry of the droplet's structure by considering the droplet as an effective dipolar moment [21,34-36]. This ellipticity of the precession as a perturbation to the system can increase the linewidth of the droplet mode, as discussed in the context of Fig. 1(c) [19]. Note that the rotational symmetry of the droplet is not broken due to the absence of drift instabilities in the system. Moreover, one can observe a small reduction of the droplet's diameter in the free layers with larger thickness. To better compare the effective diameter of the droplets, we analyze the droplet mode by performing a fast Fourier transformation on the simulated data in space and time and, mapping out the spatial distribution of the droplet mode. Under this condition, the ellipticity of the droplet's structure is averaged given to the rotation of the droplet around the normal axis and thus, we can define the effective diameter of the droplet as the distance between the precessional boundaries of the droplet. This is highlighted in Fig. 4(c), which shows the one-dimensional spatial profile of the droplet mode under the NC taken from Fig. 4(b). One can observe that the averaged droplet diameter decreases from 132 nm in a 3.5 nm thick free layer to 118 nm inside a 9.5 nm thick free layer, suggesting a better localization of the droplet under the NC. The droplet diameter dependence on the free layer thickness is summarized at the inset of Fig. 4(c), which demonstrate that the that the ratio of the NC diameter to the droplet effective diameter approached to unity in higher thicknesses.

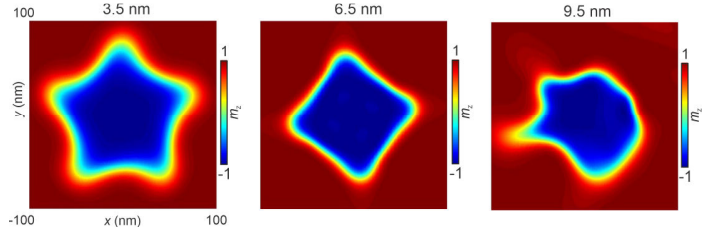

FIG.5. Snapshots of the droplet dynamics in free layers with different thickness in the presence of the Oersted field of the current.

The presence of the Oersted field of the current can lead to several instability processes of the droplet, such as drift resonances, inertial dynamics as well as the excitation of the perimeter modes [25-30]. To investigate the stability of the droplet as a function of the free layer thickness, we implement the Oersted field produced by the current in our simulations by considering the NC as an infinite wire [28]. We drive the systems with a fixed current of $I = 45$ mA. The snapshots of the droplet modes in different thicknesses are shown in Fig. 5. In addition to the droplet modes, the perimeter excitation modes (PEMs) of the droplet are excited [26]. The PEMs of the droplet is associated with the periodic spatial deformations of the droplet around the perimeter. Such modes are parametrically excited when the droplet mode is two times larger than the frequency of the perimeter modes. The results shown in Fig. 5 illustrate that the droplet in the 3.5 nm free layer loses its stability through the excitation the PEMs with the index number of $n = 5$. Interestingly, increasing the thickness to 6.5 nm leads to the excitation of the $n = 4$ PEMs. This can be explained by the smaller frequency of the droplet in thicker layers since higher PEMs indices are associated with higher frequencies [26]. In a larger thickness of 9.5 nm, the droplet loses its stability, and due to the impact of higher dipolar interactions, the phase coherency of the precession breaks, and the droplet precesses nonuniformly. This will be associated with a larger linewidth.

### D. Nucleation boundary of the droplet

The droplet is characterized by a field-current nucleation boundary [40, 41]. In the presence of a normal spin-polarizer similar to our simulations, the dependency of the nucleation boundary on the free layer thickness reads,

$$I_{throshold} = \mathcal{A} \times H_e \quad (3)$$

in which $H_e$ is the external field and $\mathcal{A}$ is a function of the saturation magnetization, NC diameter, spin torque asymmetry, electron charge, vacuum permeability, reduced Planck's constant, the spin-torque efficiency, and the free layer thickness $d$. In the systems we investigate here, only the thickness $d$ varies. Therefore, one can find the following simple ratio between two systems with different thicknesses,

$$\mathcal{A}_2/\mathcal{A}_1 = d_2/d_1 \qquad (4)$$

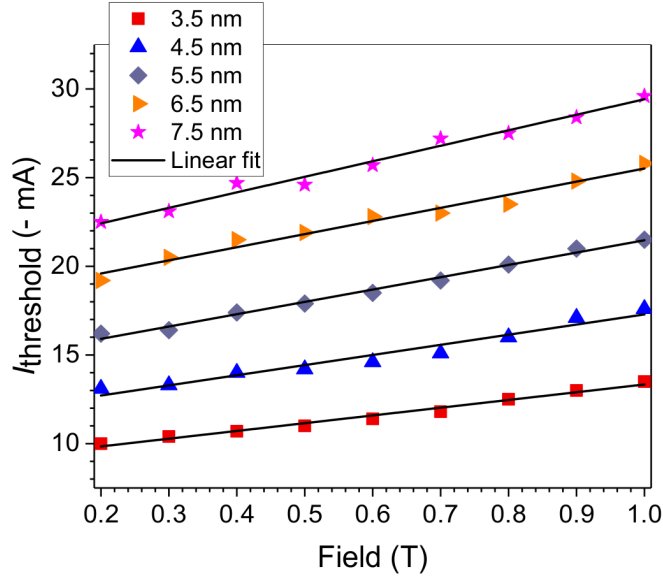

FIG.6. the droplet nucleation boundary in free layers with different thicknesses. Separated data sets are the results from simulations while the continuous black lines are linear fit to the data sets.

Figure 6 shows the nucleation boundary of the droplet in free layers with different thicknesses. As discussed before, the droplet threshold current increases via increasing the thickness. This is valid for the entire ranges of the external field that the droplet nucleates. A linear function based on Eq. (3) fits very well to the simulated data sets as shown by black lines in Fig. 6. In principle, a linear dependence of the $\mathcal{A}$ on the thickness can be observed. For example, by using a linear fit, we find $\mathcal{A}$ (3.5 nm) = 4.36 ± 0.21 (-mA/T) and, based on Eq. (4), we expect that $\mathcal{A}$ (4.5 nm) = 5.60 (-mA/T). This value is in agreement from the fit to the simulated data which is $\mathcal{A}$ (4.5nm) = 5.69 (-mA/T). We summarize the obtained results from the fitting function in table 1.

| Thickness $d$ (nm) | $\mathcal{A}$ (-mA/T) | $d / d$ (3.5 nm) | $\mathcal{A} / \mathcal{A}$ (3.5 nm) |
|---|---|---|---|
| 3.5 | 4.36 ± 0.21 | 1 | 1 |
| 4.5 | 5.69 ± 0.18 | 1.28 | 1.30 |
| 5.5 | 6.95 ± 0.34 | 1.57 | 1.59 |
| 6.5 | 7.4 ± 0.45 | 1.85 | 1.69 |
| 7.5 | 8.98 ± 0.16 | 2.14 | 2.06 |

Table 1: Summarized results of the fitting constant $\mathcal{A}$ for free layers with different thicknesses. Two right columns show the normalized values to the reference device with 3.5 nm thick free layer.

The two right columns of Table 1 show the normalized values of *d* and $\mathcal{A}$ to the reference system with a 3.5 nm free layer. Comparing the two right columns suggests a good accuracy of Eq. (4) to predict the droplet nucleation boundaries for systems whose free layers have different thicknesses.

I. **Conclusion**

In conclusion, we have investigated the impact of the dipolar interactions and the demagnetizing field on the dynamics of magnetic droplet solitons in spin-torque nano-oscillators. To this end, we have tuned the internal field of the system by varying the thickness of the free layer. Our results demonstrate that the droplet requires a higher current to nucleates in thicker films. Moreover, the droplet features a larger current hysteresis loop, a smaller precessional frequency and a semi-elliptical shape with a smaller effective diameter in thicker films. In addition, the presence of the Oersted field of the current leads to stronger droplet instabilities in free layers with larger thickness. The droplet nucleation boundary strongly depends on the thickness of the free layer, nevertheless, a simple relation was used to elaborate the nucleation boundary in thicker films, which can be employed in experimental investigations. Our study uncovers an additional degree of influence on the dynamics of magnetic droplet solitons, and further enlighten the suitability of ultrathin layers for using droplets in spintronic applications. Moreover, it motivates further investigations on the nonlinear SW dynamics in magnetic layers with large PMA.

The authors thank Burkard Hillebrands and Philipp Pirro for support and valuable discussions. This project is funded by the Deutsche Forschungsgemeinschaft (DFG, German Research Foundation) - TRR 173 - 268565370 ("Spin + X", Project B01), the DFG Priority Programme "SPP2137 Skyrmionics" and the Iran Science Elites Federation (ISEF).

*Correspondence: m-mohseni@sbu.ac.ir

References.